\documentclass[eqsecnum,twocolumn,showpacs,aps]{revtex4}
\usepackage{graphicx}
\begin{document}
\title{Dissipative hydrodynamics in 2+1 dimension}
\author{A. K. Chaudhuri}
\email[E-mail:]{akc@veccal.ernet.in}
\affiliation{Variable Energy Cyclotron Centre, 1/AF, Bidhan Nagar, 
Kolkata 700~064, India}

\begin{abstract}
In 2+1 dimension, we have simulated the hydrodynamic evolution of 
QGP fluid with dissipation due to shear viscosity. Comparison of 
evolution of ideal and viscous fluid, both initialised
under the same conditions e.g. same equilibration time, energy density
and velocity profile, reveal that the dissipative fluid evolves slowly, 
cooling at a slower rate. Cooling get still slower for higher viscosity. 
The fluid velocities on  the otherhand evolve faster in a dissipative fluid 
than in an ideal fluid.  The transverse expansion is also enhanced in dissipative evolution.
For the same decoupling temperature, freeze-out surface for a dissipative fluid is more extended
than an ideal fluid.
Dissipation produces entropy as a result of which particle production is increased. Particle production is
increased due to (i) extension of the freeze-out surface and (ii) change of the equilibrium distribution
function to a non-equilibrium one, the last effect being prominent at large transverse momentum. Compared
to ideal fluid, transverse momentum distribution of pion production is considerably enhanced. Enhancement
is more at high $p_T$ than at low $p_T$. Pion production also increases with viscosity, larger the viscosity,
more is the pion production. Dissipation also modifies the elliptic flow. Elliptic flow is reduced in viscous
dynamics. Also, contrary to ideal dynamics where elliptic flow continues to increase with transverse momentum,
in viscous dynamics, elliptic flow tends to saturate at large
transverse momentum. The analysis suggest that initial conditions of the 
hot, dense matter produced in Au+Au collisions at RHIC, as extracted from
ideal fluid analysis can be changed significantly if the QGP fluid is viscous.  
\end{abstract}


\date{\today}  

\maketitle

\section{Introduction}
 
Lattice  QCD predicts that under certain conditions (sufficiently
high energy density and temperature),  ordinary  hadronic  matter
(where  quarks  and  gluons  are  confined),  can undergo a phase
transition to a deconfined matter, commonly known as Quark  Gluon
Plasma (QGP).  Nuclear physicists are trying to produce and detect
this  new  phase  of matter at RHIC, BNL.  Recent Au+Au collisions at RHIC,
indicate that dense, color opaque medium of deconfined matter is created in 
very central collisions
\cite{BRAHMSwhitepaper,PHOBOSwhitepaper,PHENIXwhitepaper,
STARwhitepaper}.  It is now also understood that the quarks and gluons strongly interact, giving rise to  the notion of sQGP
(strongly interacting QGP).
The experimental data are successfully analysed in a {\em ideal} fluid dynamic model \cite{QGP3}. Hydrodynamic evolution
 of ideal QGP, thermalised at $\tau_i$=0.6 fm, with central entropy density of
110 $fm^{-3}$, or energy density of 35 $GeV/fm^3$ can explain a large volume of RHIC data, the $p_T$ spectra of identified
particles, the elliptic flow etc \cite{QGP3}. However, experimental data do show deviation from ideal behavior.  The ideal
fluid description works well in almost central Au+Au collisions near  mid-rapidity at top RHIC energy, but gradually breaks
down in more  peripheral collisions, at forward rapidity, or at lower collision energies \cite{Heinz:2004ar}, indicating the
onset of dissipative effects. To describe such deviations from ideal fluid dynamics quantitatively,  requires the  numerical
implementation of {\em dissipative} relativistic fluid dynamics.

Eckart \cite{Eckart} and Landau and Lifshitz \cite{LL63} formulated the theory  of dissipative relativistic fluid. Their
theories are called first order theories and suffer from the problem of causality; the signal can travel faster than light.
1st order theories  assume that the entropy 4-current contains terms upto linear order in dissipative quantities. This
restriction results in  parabolic equations, which leads to causality problem.   The causality problem is removed in the
2nd-order theories, formulated
by Israel and Stewart \cite{IS79}. In 2nd order theories,  dissipative fluxes are treated as (extended) thermodynamic
variables and the linear relation
between entropy 4-current and dissipative quantities is extended to include 
quadratic terms. The entropy 4-current contains terms upto 2nd order in 
dissipative forces. The resulting equations are hyperbolic in nature and the causality problem is removed. Naturally 2nd
order theories are more complicated. In addition to the usual energy-momentum conservations equations, relaxation equations
for dissipative fluxes are required to be
solved simultaneously. 
 
Though the theories of dissipative hydrodynamics \cite{Eckart,LL63,IS79} 
has been known for more than 30 years , significant progress towards its numerical implementation  has only been made very
recently \cite{Muronga:2001zk,Teaney:2004qa,MR04,CH05,Heinz:2005bw}.
Earlier attempts were restricted to simple one-dimensional Bjorken 
expansion \cite{vis_bj}. Since the hot dense matter, produced in RHIC collisions, undergoes large transverse expansion,
such models are not
of much help in extracting initial conditions of the fluid from the experiment.
Recently Teaney \cite{Teaney:2004qa} solved the hydrodynamic equations in a 1st order dissipative theory. He used the blast
wave model and calculated the corrections to thermal distribution functions considering shear viscosity only. Blast wave
models lack dynamics. Freeze-out surface is parameterised and one fit the freeze-out parameters with experimental observables.
 Information about the initial conditions of the hot dense matter is not available in blast wave models. In \cite{MR04,CH05},
 2nd
order theories for dissipative fluid, is  solved for QGP fluid. While the models
 are dynamic, (energy-momentum conservation equations are solved)
 energy-momentum conservation equations  are solved in 1+1 dimension,
 assuming longitudinal  boost-invariance and cylindrical symmetry. Thus
 these models do not give any information on the most sensitive experimental
 observable, the elliptic flow.  Recently in \cite{Heinz:2005bw} explicit
 equations describing the space-time evolution of non-ideal relativistic
 fluid, undergoing boost-invariant  longitudinal and arbitrary transverse
 expansion, are given. Equations are written for both the 1st order and
 2nd order theory.

At the Variable Energy Cyclotron Centre, Kolkata, we have developed a  numerical code  (AZHYDRO-KOLKATA) to solve, both the 1st
order and 2nd order dissipative hydrodynamics in 2+1 dimension.  In the present paper we will present  AZHYDRO-KOLKATA results
for  the 1st order dissipative hydrodynamics. Results for 2nd order dissipative hydrodynamics will be presented in a later
publication.
As mentioned earlier, 1st order theories are acausal, signal can travel faster
than light.   In one dimension, 1st order theories can be solved analytically. Analytical solutions indicate that at early
time, an unphysical reheating of the fluid can occur \cite{MR04,Baier:2006um}. Even though
in the present simulation, we donot find any indication of early reheating,
nevertheless, we maintain that only 2nd order results,  which do not have the unphysical causality the problem are more
reliable.
In this paper we will be concerned mainly with the effect of viscosity on fluid evolution and its affects the particle
production; $p_T$ distribution and elliptic flow. No attempt will be  made to explain experimental data.

Paper is organised as follows: in section II, we describe briefly the
hydrodynamic equations needed for 1st order dissipative hydro-dynamics.
In section III, we describe the equation of state, shear viscosity
coefficient and initial conditions used in the present study. With dissipation,
equilibrium distribution function is changed. Correction to equilibrium
distribution function due to non-equilibrium effects are described in 
section IV. We have studied particle (pion) production in the model. The relevant equations for pion production with
the equilibrium distribution
function and its correction due to non-equilibrium effects, are discussed in section V. Results of the numerical
simulations are shown in
section VI.   Lastly, in section VII, summary and conclusions are drawn.

\section{1st order dissipative fluid dynamics}
 
Relativistic dissipative hydrodynamics and associated equations
in 2+1 dimensions has been discussed in detail in ref.\cite{Heinz:2005bw}. 
Any fluid dynamical approach starts from the
conservation laws for the conserved charges and for energy-momentum.
For a singly charged fluid, the conservation laws are:
\begin{eqnarray}
\label{eq1}
\partial_\mu N^\mu &=&0\\
\label{eq2}
\partial_\mu T^{\mu\nu} &=& 0.
\end{eqnarray}
It must also ensure the second law of thermodynamics
\begin{equation}
\label{eq3}
\partial_\mu S^\mu \geq 0,
\end{equation}
where $S^\mu$ is the entropy current. 

In the present paper we restrict ourselves to central rapidity region,
where the QGP fluid is essentially baryon free. We thus neglect 
Eq.\ref{eq1}. To keep the calculations simple, we consider the most
important dissipative term, the shear viscosity
and neglect the other dissipative terms, e.g. heat conduction, bulk viscosity.
For a baryon free fluid, the heat conduction is zero. Bulk
viscosity is zero for QGP (point particles). Bulk viscosity will
be non-zero in the hadronic phase but presently we neglect it.

With the help of hydrodynamic 4-velocity $u^\mu$ (normalised as
$u^\mu u_\mu=1$) and projector $\Delta^{\mu\nu}=g^{\mu\nu}-u^\mu u^\nu$,
with only shear viscosity as the dissipative flux, in Landau's frame, 
 the energy momentum tensor, 
and the entropy 4-current, can be decomposed as,

\begin{eqnarray}
\label{eq4}
 &&T^{\mu\nu} = T_{\rm eq}^{\mu\nu} + \delta T^{\mu\nu}
           = \varepsilon\,u^\mu u^\nu -p\Delta^{\mu\nu}
               + \pi^{\mu\nu} \\
\label{eq5}
&&S^\mu = S^\mu_{\rm eq} +\delta S^\mu = s\,u^\mu + \Phi^\mu.
\end{eqnarray}
 
\noindent where    
$\varepsilon=u_\mu T^{\mu\nu} u_\nu$ is the energy density, $p$ is the local
pressure. $\pi^{\mu\nu}$ is the non-ideal part of the energy-momentum
tensor, the    stress tensor due to shear viscosity.   
$s=u_\mu S^\mu$ is the entropy density and $\Phi^\mu$ is the entropy
flux due to dissipation.
 
In the first order theories, the shear stress tensors are written as,
 
\begin{equation}
\label{eq6}
\pi^{\mu\nu} = 2\eta  \nabla^{<\mu}u^{\nu>}
\end{equation}

\noindent where $\eta$ is the shear viscosity coefficient and 
$\nabla^{<\mu}u^{\nu>}$ is a trace-less symmetric tensor defined as,

\begin{equation} \label{7}
 \nabla^{<\mu}u^{\nu>}=
  \left[\frac{1}{2}\left(\Delta^{\mu\sigma}
\Delta^{\nu\tau}{+}\Delta^{\nu\sigma}\Delta^{\mu\tau}\right)-\frac{1}{3}
\Delta^{\mu\nu}\Delta^{\sigma\tau}\right] 
\end{equation}

Viscous pressure $\pi^{\mu\nu}$  is symmetric ($\pi^{\mu\nu}=\pi^{\nu\mu}$),
traceless ($\pi^\mu_\mu=0$) and transverse to hydrodynamic velocity,
($u_\mu \pi^{\mu\nu}=0)$. The 16-component $\pi^{\mu\nu}$ has only 5
independent components. As mentioned in the beginning, we have solved the
equations with the assumption of longitudinal boost-invariance. With boost-invariance
the number of independent shear-stress tensors further reduces to 3. 

Heavy ion collisions are best described in terms of proper time $\tau=\sqrt{t^2-z^2}$ and
rapidity $\eta_s=\frac{1}{2}\ln \frac{t+z}{t-z}$ (we use the subscript $s$ to distinguish spatial rapidity from the
viscous coefficient $\eta$). 
In $(\tau,x,y,\eta_s)$ coordinates, with longitudinal boost-invariance,
the hydrodynamic 4-velocity can be written as, 

\begin{eqnarray}
u^\mu=&&(u^\tau,u^x,u^y,u^{\eta_s}) \nonumber\\
          =&&(\gamma_\perp,\gamma_\perp v_x, \gamma_\perp v_y,0),
\end{eqnarray}

\noindent with $\gamma_\perp=1/\sqrt{1-v_x^2-v_y^2}$.
Explicit equations for energy-momentum conservation,
in $(\tau,x,y,\eta_s)$ coordinate system has been developed in ref.\cite{Heinz:2005bw}. 
Here we rewrite the results in a form suitable for numerical algorithm.
The 
energy-momentum conservation equations are,

\begin{widetext}
\label{eq8}
\begin{eqnarray}
\label{eq8a}
&& \partial_\tau\Bigl(\tilde{T}^{\tau\tau}\Bigr) 
 +\partial_x \Bigl(\tilde{T}^{\tau \tau} \overline{v}_x \Bigr)+\partial_y \Bigl(\tilde{T}^{\tau \tau} \overline{v}_y \Bigr)= 
  -\,(p+\tau^2 \pi^{\eta\eta}),
\\
\label{eq8b}
&&\partial_\tau\Bigl(\tilde{T}^{\tau x}\Bigr) 
 +\partial_x \Bigl(\tilde{T}^{\tau x}v_x\Bigr)
 +\partial_y \Bigl(\tilde{T}^{\tau x}v_y\Bigr) 
 = -\partial_x(\tilde{p} + \tilde{\pi}^{xx}-\tilde{\pi}^{\tau x} v_x) - \partial_y(\tilde{\pi}^{xy}-\tilde{\pi}^{\tau x}v_y),
\\
\label{eq8c}
&&\partial_\tau\Bigl(\tilde{T}^{\tau y}\Bigr) 
 +\partial_x \Bigl(\tilde{T}^{\tau y}v_x\Bigr)
 +\partial_y \Bigl(\tilde{T}^{\tau y}v_y\Bigr) 
 = -\partial_x(\tilde{\pi}^{xy}-\tilde{\pi}^{\tau y} v_x) - \partial_y(\tilde{p} + \tilde{\pi}^{yy}-\tilde{\pi}^{\tau y}v_y),
\end{eqnarray}

\noindent where $\overline{v}_x=T^{\tau x}/T^{\tau\tau}$ and $\overline{v}_y=T^{\tau y}/T^{\tau\tau}$, and we have used
the notation
"tilde" to represent quantities multiplied by the factor $\tau$, 
$\tilde{p}=\tau p$ and similarly   $\tilde{T}^{ij}=\tau T^{ij}$. 
We note that unlike in ideal fluid, in viscous fluid dynamics 
conservation equations contain additional pressure gradients containing
the dissipative fluxes.  
Both $T^{\tau x}$ and $T^{\tau y}$ components of energy-momentum
tensors now evolve under
the influence of additional pressure gradients.  

In 1st order theory, the shear stress tensor components required  in the 
preceding equations are,

\begin{eqnarray} \label{9}
&&\pi^{\tau x} =2\eta[ - \frac{1}{2}\partial_x\gamma_\perp 
                    + \frac{1}{2}\partial_\tau(\gamma_\perp v_x)
                    - \frac{1}{2} D(\gamma_\perp^2 v_x)
                    + \frac{\theta}{3}\,\gamma_\perp^2v_x],
\\
&&\pi^{\tau x} =2\eta[ - \frac{1}{2}\partial_y\gamma_\perp 
                    + \frac{1}{2}\partial_\tau(\gamma_\perp v_y)
                    - \frac{1}{2} D(\gamma_\perp^2 v_y)
                    + \frac{\theta}{3}\,\gamma_\perp^2v_y],\\
&&\pi^{\tau\tau} =2\eta[ \frac{\theta}{3}(\gamma_\perp^2-1) 
  + \partial_\tau\gamma_\perp-\frac{1}{2}D(\gamma_\perp^2)],
\\
&&\pi^{\eta\eta} =2\eta[ \frac{1}{\tau^2}
  \left(\frac{\theta}{3}-\frac{\gamma_\perp}{\tau}\right)],
\\
&&\pi^{xx}=2\eta[-\partial_x(\gamma_\perp v_x)-\frac{1}{2}D(\gamma_\perp^2)+
 \frac{\theta}{3}(1+\gamma_\perp^2 v_x^2)], \\
&&\pi^{yy}=2\eta[-\partial_x(\gamma_\perp v_y)-\frac{1}{2}D(\gamma_\perp^2)+
 \frac{\theta}{3}(1+\gamma_\perp^2 v_y^2) ]
\end{eqnarray}
\end{widetext}

\noindent where $D=u^\mu \partial_\mu$ is the convective time derivative,

\begin{equation}
D=\gamma_\perp(\partial_\tau + v_x \partial_x + v_y \partial_y),
\end{equation}

\noindent and $\theta$ is the  local expansion rate, given by,

\begin{equation}
\theta=\frac{\gamma_\perp}{\tau}+\partial_\tau \gamma_\perp +
\partial_x(v_x\gamma_\perp)+\partial_y(v_y \gamma_\perp)
\end{equation}
 

Given an equation of state, if
energy density ($\varepsilon$) and fluid velocity ($v_x$ and $v_y$) distributions, at any time
$\tau_i$ are known, Eqs.\ref{eq8a},\ref{eq8b} and \ref{eq8c} can be  integrated
to obtain $\varepsilon$,$v_x$ and $v_y$ at the next time step $\tau_{i+1}$. 
While for ideal hydrodynamics, this procedure works perfectly,   
  viscous hydrodynamics poses a problem
that shear stress-tensor components contains 
time derivatives, $\partial_\tau \gamma_\perp$, $\partial_\tau u^x$,
$\partial_\tau u^x$ etc. Thus at time step $\tau_i$ one needs the  
still unknown time derivatives. 
In 1st order theories, this problem is circumvented by calculating the
time derivatives from the ideal equation of motion,

\begin{eqnarray}
\label{eq220}
Du^\mu &=&\frac{\nabla^\mu p}{\varepsilon+p},\\
\label{eq221}
D\varepsilon&=&-(\varepsilon+p)\nabla_\mu u^\mu.
\end{eqnarray}

With the help of these two equations all the time derivatives can be 
expressed entirely in terms of spatial gradients. 
1st order theories are restricted to contain terms at most linear
in dissipative quantities.  Neglect of viscous terms can contribute only
in 2nd order corrections, which is neglected in 1st order theories.

\section{Equation of state, viscosity coefficient and initial conditions}

\subsection{Equation of state}

One of the most important inputs of a hydrodynamic model is  
the equation of state. Through this input, the macroscopic hydrodynamic models make contact 
with the microscopic world. 
In the present calculation we have used the equation of state,
EOS-Q, developed in ref.\cite{QGP3}.  It is a two-phase equation of
state. The hadronic phase of EOS-Q is modeled
as a non-interacting gas of hadronic resonance. As the temperature is increased,
larger and larger fraction of available energy goes into production of heavier and heavier resonances. This results into a
soft equation of state,
with small speed of sound, $c^2_s \approx 0.15$. With increasing temperature, the available volume is filled up with
resonances and the
hadronic states starts to overlap, and microscopic degrees of freedom
are changed from   hadron to deconfined quarks and gluons. The
QGP phase  is modeled as that of a 
non-interacting quark (u,d and s) and gluons, confined by a bag pressure
B. Corresponding equation of state, $p=\frac{1}{3}e -\frac{4}{3}B$ is 
stiff with a speed of sound $c_s^2 =\frac{1}{3}$. The two phases 
are matched by Maxwell construction at the critical temperature, $T_c=164 MeV$, adjusting the 
Bag pressure $B^{1/4}$=230 MeV. As discussed in \cite{QGP3}, ideal
hydrodynamics explain a large volume of RHIC Au+Au 
data with EOS-Q.

\subsection{Shear viscosity coefficient}

Shear viscosity coefficient ($\eta$) of dense nuclear (QGP or 
resonance hadron gas) is quite uncertain.
In perturbative regime, shear viscosity of a QGP is
estimated \cite{Arnold:2000dr,Baym:1990uj},

\begin{equation}
\eta=86.473 \frac{1}{g^4}\frac{T^3}{log g^{-1}},
\end{equation}

With entropy of QGP, $s=37\frac{\pi^2}{15}T^3$ and 
$\alpha_s \approx$0.5, the ratio of viscosity over the entropy, in the perturbative regime is estimated as,

\begin{equation}
\left (\frac{\eta}{s} \right )_{pert} \approx 0.135,
\end{equation}

However, QGP  produced in nuclear collisions is non-perturbative. It is
 strongly interacting QGP. Recently, using the ADS/CFT correspondence
\cite{Policastro:2001yc,Policastro:2002se},
shear viscosity of a strongly coupled gauze theory, N=4 SUSY YM,
has been evaluated, $\eta=\frac{\pi}{8}N^2_cT^3$ and the    entropy
is given by $s=\frac{\pi^2}{2}N^2_cT^3$. Thus in the strongly coupled field theory,

\begin{equation}
\left ( \frac{\eta}{s} \right )_{ADS/CFT} = \frac{1}{4\pi}\approx0.08,
\end{equation}

\noindent  
which  is 2 times larger than the perturbative estimate. In the
present paper, we treat the shear viscosity as a parameter of the
model. 
To demonstrate the effect of viscosity on flow and subsequent particle
 production, we use both the perturbative and ADS/CFT estimate of viscosity.

1st order theories are acuasal. As mentioned earlier, unphysical effects
like reheating of the fluid, early in the evolution, can occur. In one dimension
energy-momentum conservation equation can be solved analytically.
If initial fluid temperature is $T_i$ at initial time $\tau_i$, for constant 
$\eta/s$, the fluid temperature at time $\tau$ can be obtained as \cite{Baier:2006um},

\begin{equation}
T(\tau)=T_i \left (\frac{\tau_i}{\tau} \right )^{1/3} \left [1+
\frac{2}{3 \tau_i T_i} \frac{\eta}{s} \left(1- \left(\frac{\tau_i}{\tau}\right)^{2/3}\right)
\right ]
\end{equation}

For early times, $\tau < \tau_{max}$,

\begin{equation}
\tau_{max}=\tau_i\left (\frac{1}{3}+\frac{s}{\eta}
\frac{\tau_i T_i}{2}\right)^{-3/2},
\end{equation}

\noindent the solution shows an unphysical reheating. The unphysical reheating is minimised if $\tau_{max}$ is small or
$\eta/s << \tau_i T_i$.
As will be explained below, we have used an initial time $\tau_i$=0.6 
fm/c and initial temperature of the fluid, $T_i$=0.35 GeV . For both the values of viscosity, $\eta/s << \tau_i T_i$, the
unphysical reheating is minimised. 

Shear viscosity can also be expressed in terms of sound attenuation
length, $\Gamma_s$, defined as,

\begin{equation}
\Gamma_s=\frac{\frac{4\eta}{3}}{sT}
\end{equation}

$\Gamma_s$ is equivalent to mean free path and for a 
valid hydrodynamic description $\Gamma_s/\tau << 1$, i.e. mean
free path is much less than the system size.  With the present choice of
equilibration time and temperature, both for ADS/CFT and perurbative estimate of viscosity,
at initial time, $\Gamma_s/\tau$ is much less
than unity and hydrodynamics remains a valid description.
At later time the validity
condition becomes even better. 

\subsection{Initial conditions}

As discussed earlier, ideal hydrodynamics has been very successful
in explain a large volume of data in RHIC 200AGeV Au+Au collisions 
 \cite{QGP3}. 
In the present demonstrative calculations,  we have used the 
similar initial conditions as in ref.\cite{QGP3}. 
Details of the initial conditions can be found
in \cite{QGP3}. We just mention that
in ref.\cite{QGP3}, initial
transverse energy is parameterised geometrically.
At an impact parameter $\vec{b}$, transverse distribution of
wounded nucleons $N_{WN}(x,y,\vec{b})$    and of binary NN collisions 
$N_{BC}(x,y,\vec{b})$ to are calculated in a Glauber model. 
A collision at impact parameter $\vec{b}$ is assumed to contain 25\% hard scattering
(proportional to number of binary collisions) and 75\% soft scattering
(proportional to number of wounded nucleons).  
Transverse energy density 
profile at impact parameter $\vec{b}$ is then obtained as,

\begin{equation}
\varepsilon(x,y,\vec{b})=\varepsilon_0(0.75\times N_{WN}(x,y,\vec{b})+0.25 \times N_{BC}(x,y,\vec{b}))
\end{equation}

The parameter $\varepsilon_0$ and the initial equilibration time
$\tau_i$ are fixed to reproduce the 
experimental transverse momentum distribution of pions in central Au+Au collisions. 
STAR and PHENIX data are fitted to obtain initial
equilibrium time $\tau_i$=0.6 fm and central  entropy density of $s=110 fm^{-3}$. 
This corresponds energy density of the
fluid as 25 $GeV/fm^3$, or 
initial temperature of 350 MeV.  Apart from initial energy density, 
initial velocity distribution is also required in hydrodynamic
calculation. In the present
calculation it is assumed that the at the initial time $\tau_i$, fluid
velocities are zero, $v_x(x,y)=v_y(x,y)=0$.
 
In dissipative hydrodynamics, additionally, initial conditions 
for the dissipative fluxes needs to be specified. In the present paper 
we assume that by the equilibration time $\tau_i$, the dissipative fluxes
attained their longitudinal  boost-invariant values. 
 
\begin{eqnarray}
\pi^{\tau x} =&&0\\
\pi^{\tau x} = &&0\\
\pi^{\tau\tau} =&&0\\
\tau^2 \pi^{\eta\eta} =&&-4\eta/\tau_i \\ 
\pi^{xx}=&&2\eta/\tau_i\\ 
\pi^{yy}=&&2\eta/\tau_i 
\end{eqnarray}

\section{Non-equilibrium distribution function}

With dissipation the system is not in equilibrium and the equilibrium 
distribution function,

\begin{equation}
f^{(0)}(x,p)=\frac{1}{exp[\beta(u_\mu p^\mu -\mu)] \pm 1},
\end{equation} 

\noindent with inverse temperature $\beta=1/T$ and chemical potential
$\mu$
 can no longer describe the system. In a highly
non-equilibrium system,   distribution
function is unknown. If the system is slightly off-equilibrium,  then
it is possible to calculate correction to equilibrium distribution 
function due to   (small) non-equilibrium effects. Slightly
off-equilibrium distribution function can be  approximated  as, 

\begin{equation}
F(x,p)=f^{(0)}(x,p) [1+\phi(x,p)],
\end{equation}

\noindent  $\phi(x,p)$ is the deviation from equilibrium distribution 
function  $f^{(0)}$. With shear viscosity as the only dissipative forces,
$\phi(x,p)$ can be locally approximated by a quadratic function 
of 4-momentum,

\begin{equation}
\phi(x,p)=\varepsilon_{\mu\nu} p^\mu p^\nu.
\end{equation}

Without any loss of generality $\varepsilon_{\mu\nu}$ can be written as
as,

\begin{equation} \label{eq4_4}
\varepsilon_{\mu\nu}=C \pi^{\mu\nu}, C=\frac{\beta^2}{2(\varepsilon+p)},
\end{equation}

\noindent completely specifying the non-equilibrium distribution function.

\section{particle spectra}

With the non-equilibrium distribution function thus specified, it can be used to 
calculate the particle spectra from the freeze-out surface. In the standard 
Cooper-Frye
prescription, particle distribution is obtained as,

\begin{equation} \label{eq5_1}
E\frac{dN}{d^3p}=\frac{dN}{dyd^2p_T} =\int_\Sigma d\Sigma_\mu p^\mu f(x,p)
\end{equation}

In $(\tau,x,y,\eta_s)$ coordinate, the freeze-out surface is parameterised as,

\begin{equation}
\Sigma^\mu=(\tau_f(x,y)\cosh \eta_s, x, y, \tau_f(x,y) \sinh \eta_s),
\end{equation}

\noindent and the normal vector on the hyper surface is,

\begin{equation}
d\Sigma_\mu=(\cosh \eta_s, -\frac{\partial \tau_f}{\partial x_f}, 
                        -\frac{\partial \tau_f}{\partial y_f}, -\sinh \eta_s)
\tau_f dx dy d\eta_s
\end{equation}

At the fluid position $(\tau,x,y,\eta_s)$ the particle 4-momenta are parameterised as,

\begin{equation}
p^\mu=(m_T cosh (\eta_s-Y), p^x, p^y, m_T sinh (\eta_s-Y))
\end{equation}

The volume element $p^\mu d\Sigma_\mu$ become,

\begin{equation}
p^\mu d\Sigma_\mu=(m_T cosh(\eta-Y)-\vec{p}_T. \vec{\nabla}_T \tau_f) \tau_f dx dy d\eta
\end{equation}

Equilibrium distribution function involve the term $\frac{p^\mu u_\mu}{T}$ which can be evaluated as,

\begin{equation}
\frac{p^\mu u_\mu}{T}=\frac{\gamma(m_T cosh(\eta-Y) -\vec{v}_T.\vec{p}_T -\mu/\gamma)}{T}
\end{equation}

The non-equilibrium distribution function require the sum
$p^\mu p^\nu \pi_{\mu\nu}$,

\begin{equation}
p_\mu p_\nu \pi^{\mu\nu}=a_1 cosh^2 (\eta-Y) +a_2 cosh(\eta-Y) + a_3
\end{equation}

with
\begin{eqnarray}
a_1=&&m_T^2(\pi^{\tau \tau} +\tau^2 \pi^{\eta \eta})\\
a_2=&&-2m_T(p_x \pi^{\tau x} + p_y \pi^{\tau y})\\
a_3=&&p_x^2 \pi^{xx} +p_y^2 \pi^{yy} +2p_x p_y \pi^{xy} - m_T^2 \tau^2 \pi^{\eta \eta}
\end{eqnarray}

Inserting all the relevant formulas in Eq.\ref{eq5_1} and integrating over 
spatial rapidity one obtains,
 
\begin{equation}
\frac{dN}{dyd^2p_T} =\frac{dN^{eq}}{dyd^2p_T}+\frac{dN^{neq}}{dyd^2p_T}
\end{equation}

with,

\begin{widetext}
\begin{eqnarray} \label{eq5_12}
\frac{dN^{eq}}{dyd^2p_T}=\frac{g}{(2\pi)^3} 
\int dx dy \tau_f [m_T K_1(n\beta) - p_T \vec{\nabla}_T \tau_f K_0(n\beta)]\\
\frac{dN^{neq}}{dyd^2p_T}=\frac{g}{(2\pi)^3} 
\int dx dy \tau_f [m_T\{ {\frac{a_1}{4} K_3(n\beta)+\frac{a_2}{2} K_2(n\beta)+(\frac{3a_1}{4}+a_3+1)K_1(n\beta)
+\frac{a_2}{2} K_0(n\beta)} \} \nonumber\\
- \vec{p}_T. \vec{\nabla}_T \tau_f \{\frac{a_1}{2} K_2(n\beta)+
 a_2 K_1(n\beta)+( \frac{a_1}{2}+a_3+1)K_0(n\beta)\}]
\end{eqnarray}
\end{widetext}

\noindent where  $K_0$, $K_1$, $K_2$ and $K_3$ are the modified Bessel functions.  

We will also show results for elliptic flow $v_2$. It is defined as,

\begin{equation}
V_2=\frac
{\int_0^{2\pi} \frac{dN}{dyd^2p_T} \cos(2\phi) d\phi}
{\int_0^{2\pi} \frac{dN}{dyd^2p_T}  d\phi}
\end{equation}

\begin{figure}[h]
\includegraphics[bb=14 13 581 829,width=0.99\linewidth,clip]{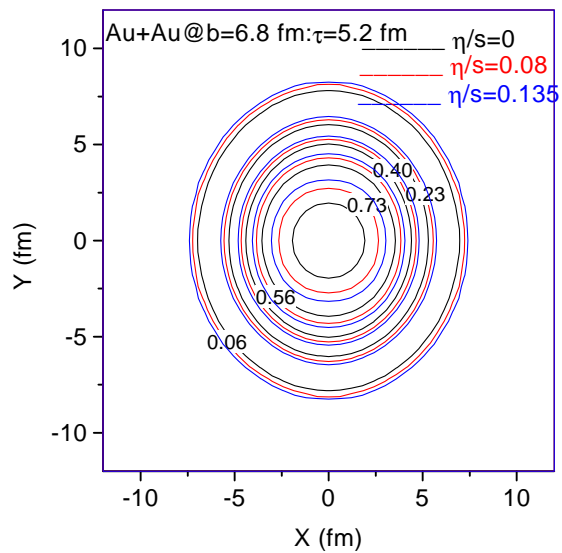}
\vspace{-4cm}
\caption{(color online). Contour plots of energy density at  (proper) time $\tau$=5.6 fm.
The black  lines are for ideal fluid ($\eta/s$=0). The red and blue
lines are for viscous fluid with ADS/CFT ($\eta/s$=0.08) and perturbative ($\eta/s$=0.135) estimate of viscosity.}
\label{F1}
\end{figure}

\section{Results}

\subsection{Evolution of the viscous fluid}

The energy-momentum conservation equations \ref{eq8a},\ref{eq8b},\ref{eq8c} are solved using the SHASTA-FCT algorithm. We
have made extensive changes to the publicly available
code AZHYDRO (described in \cite{QGP3}) for simulation of ideal fluid.
The modified code, called "AZHYDRO-KOLKATA",  simulate the evolution of dissipative fluid, both in 1st and 2nd order theory.
  In this section we present
results of AZHYDRO-KOLKATA for the 1st order theory of dissipative fluid.
In the following we will show the results obtain in Au+Au  collision at impact parameter $b=$ 6.8 fm, which approximately
corresponds
to 16-24\% centrality Au+Au collisions.  With the same initial conditions,
we have solved the energy-momentum conservation equations for
ideal fluid and viscous fluid.  

 In Fig.1, we have shown
the constant energy density contour plot in x-y plane, after an evolution of 
5 fm.
The black lines are for ideal fluid evolution. The  red and blue lines are for
viscous fluid with ADS/CFT ($\eta/s$=0.08) and perturbative ($\eta/s$=0.135) estimate of viscosity. Constant energy density
contours,
as depicted in Fig.1, indicate that with viscosity fluid cools slowly. Cooling
gets slower as viscosity increases.
Thus at any point in the x-y plane, viscous fluid temperature is higher
than that of the ideal fluid.  The results are in accordance with our
expectation. For dissipative fluid, ideal equation of motion Eq.\ref{eq221} is changed to,

\begin{equation}
\label{eq61a}
D\varepsilon=-(\varepsilon+p)\nabla_\mu u^\mu + \pi^{\mu\nu} \nabla_{<\mu} u_{\nu>}\\
\end{equation}

Due to viscosity, evolution of energy density is slowed down.   

\begin{figure}
\includegraphics[bb=14 13 581 829,width=0.99\linewidth,clip]{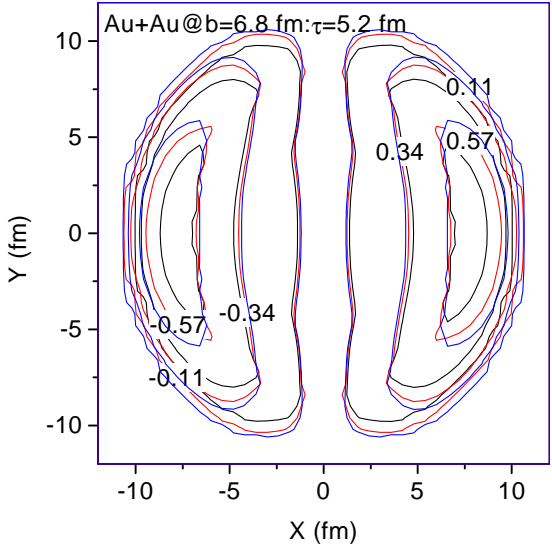}
\vspace{-4cm}
\caption{(color online). Contour plots of x-component of fluid velocity $v_x$ at 
$\tau$=5.6 fm. The black  lines are for ideal fluid ($\eta/s$=0). The 
red and blue lines are for viscous fluid with ADS/CFT and perturbative estimate of viscosity, $\eta/s$=0.08 and 0.135.}
\label{F2}
\end{figure}

In Fig.2, we have shown the constant $v_x$ contour plot in x-y plane
again at $\tau$=5.6 fm. As before the black lines
are for the ideal fluid evolution. The red and blue lines are for viscous fluid
with $\eta/s$=0.08 and 0.135 respectively. In the central region of the
fluid, viscous fluid has more velocity than its ideal counterpart. With
viscosity  while the energy density evolve slowly, the fluid velocity
evolve faster. Contour plot of the y-component of fluid velocity also
indicate similar results.

To obtain an idea of transverse expansion of viscous fluid, as opposed
to ideal fluid, in Fig.3, we have shown the constant temperature
contours in $\tau-x$ plane ,  at a fixed value of y=0 fm. Transverse expansion
is substantially enhanced in a viscous fluid. More the viscosity, more
is the transverse expansion. The plot also indicate
that at late time, fluid at x=y=0 behaves similarly to a ideal fluid.  

\begin{figure}
\includegraphics[bb=14 13 581 829,width=0.99\linewidth,clip]{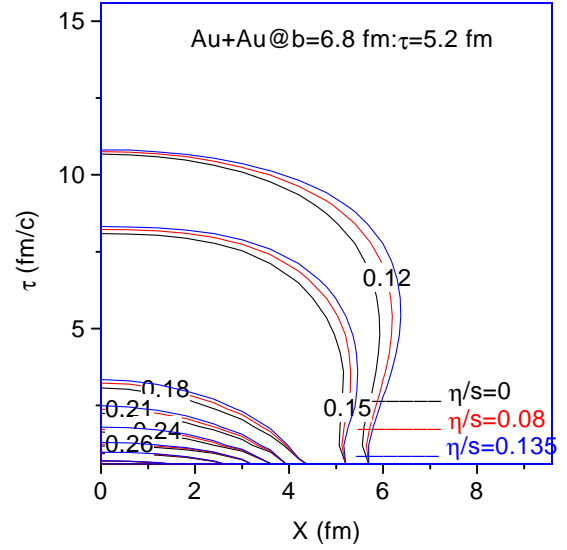}
\vspace{-4cm}
\caption{(color online). Contour plots of temperature at y=0 fm in $x-\tau$ plane. 
The black  lines are for ideal fluid ($\eta/s$=0). The red and blue
lines are for viscous fluid with $\eta/s$=0.08 and 0.135 respectively.}
\label{F3}
\end{figure}

\begin{figure}
\includegraphics[bb=14 13 581 829,width=0.99\linewidth,clip]{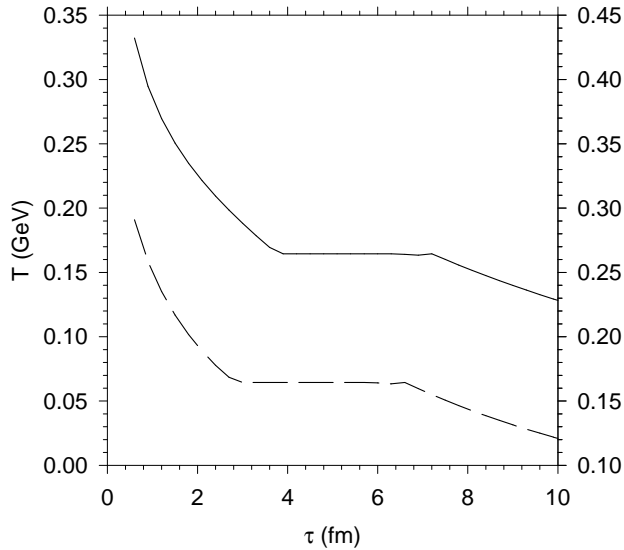}
\vspace{-4.5cm}
\caption{Evolution of the temperature in viscous dynamics with perturbative estimate of viscosity, $\eta/s$=0.135. The
solid and dashed
lines are for fluid  at x=y=0 and x=0,y=3 fm respectively. The x=0,y=3 fm curve is plotted with the right side scale.  } 
\label{F3a}
\end{figure}

1st order dissipative theories are acausal. As mentioned earlier, acausality can lead to unphysical behavior like reheating
 of the fluid in the early stage of evolution \cite{MR04,Baier:2006um}. Do we see any reheating?
In Fig.\ref{F3a}, the evolution of temperature in viscous dynamics,
with perturbative estimate of viscosity ($\eta/s$=0.135) is
shown. We have shown the temperature at two positions of the fluid,
x=y=0  (the solid line) and x=0,y=3 fm  (the dashed line).  
In both the positions of the fluid, with time as the fluid expands, temperature  
decreases (as it should be). We find no evidence of reheating. Reheating is not seen also  
with ADS/CFT estimate of viscosity.

\begin{figure}
\includegraphics[bb=14 13 581 829,width=0.99\linewidth,clip]{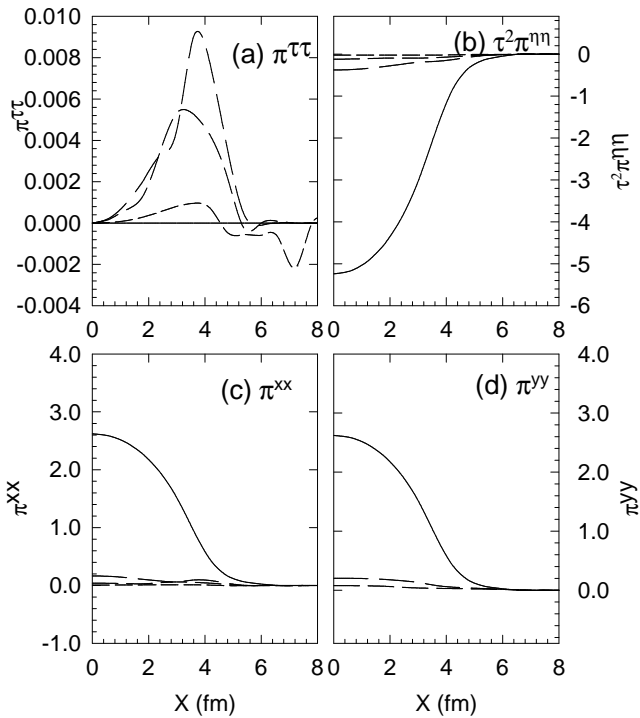}
\vspace{-2cm}
\caption{Shear stress tensor  $\pi^{\tau\tau}(x,y=0)$, $\tau^2\pi^{\eta\eta}(x,y=0)$,
$\pi^{xx}(x,y=0)$ and $\pi^{yy}(x,y=0)$ at 
$\tau$=0.6, 2.2, 3.2 and 4.2   fm are shown in 4 panels.
 }
\label{F4}
\end{figure}

In Fig.\ref{F4}, in 4-panels we have shown    shear stress tensors $\pi^{\tau\tau}(x,y=0)$, $\tau^2 \pi^{\eta\eta}(x,y=0)$,
$\pi^{xx}(x,y=0)$ and $\pi^{yy}(x,y=0)$ as a function of x.
$\eta/s$ =0.135.
The solid, long-dashed, dashed and short-dashed lines
are for time 0.6, 2.2, 3.2 and 4.2 fm respectively.
Initially at $\tau$=0.6 fm, $\pi^{\tau\tau}$ is zero. As the fluid evolve, $\pi^{\tau\tau}$ increases rapidly to a maximum
and
 then decreases. By 4 fm of evolution, 
it decreases to very small values. We also note that $\pi^{\tau\tau}$
is never very large. The viscous pressures
 $\tau^2 \pi^{\eta\eta}$, $\pi^{xx}$
and $\pi^{yy}$ are non-zero at initial time $\tau_i$=0.6 fm
As the fluid evolve these viscous fluxes rapidly decreases to
very small values.

Viscosity generates entropy. In the model entropy generation
due to dissipation can be calculated as,

\begin{equation}
\partial_\mu S^\mu = \frac{\pi^{\mu\nu} \pi_{\mu\nu}}{2\eta T} 
\end{equation}

\begin{figure}
\includegraphics[bb=14 13 581 829,width=0.99\linewidth,clip]{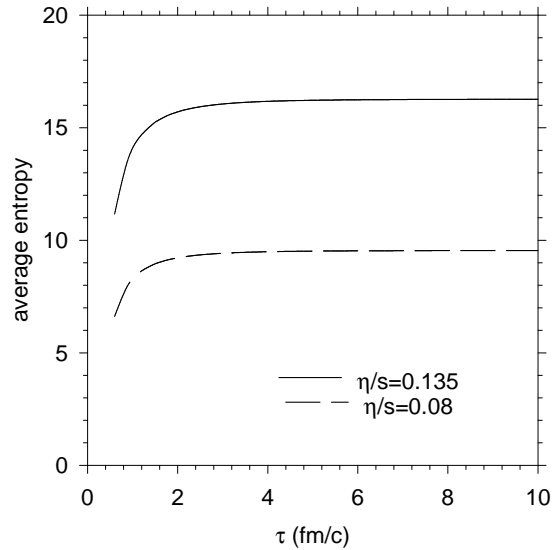}
\vspace{-4.5cm}
\caption{Evolution of average entropy with proper time,
for two values of $\eta/s$ is shown.}
\label{F5}
\end{figure}

Evolution of spatially average entropy is shown in Fig.\ref{F5}.   Entropy generation saturates after $\sim$ 2 fm of evolution.
It is expected.
As seen in Fig.\ref{F4}, viscous fluxes rapidly decreases and
by 2 fm of evolution, the viscous fluxes are decreased 
sufficiently
and  do not contribute significantly to the entropy.

\subsection{Particle spectra}

In this exploratory calculations we have not attempted to fit experimental
data. We just exhibit the effect of viscosity on (i) transverse momentum
distribution and (ii) elliptic flow of pions.   Viscosity
influences the particle production by (i) changing the freeze-out surface (freeze-out surface is extended) and (ii) by introducing
a correction to the
equilibrium distribution function.  
Non-equilibrium correction to equilibrium distribution function
  depend, quadratically on the momentum and linearly on the 
viscous fluxes. 

\begin{figure}
\includegraphics[bb=14 13 581 829,width=0.99\linewidth,clip]{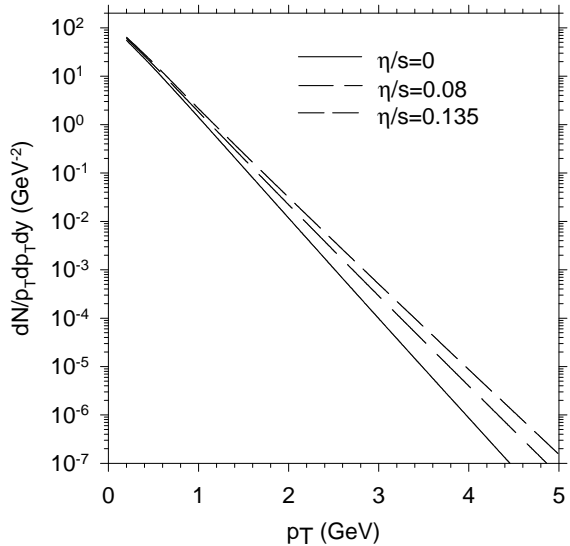}
\vspace{-4.5cm}
\caption{$P_T$ distribution of pions. The solid line is for ideal fluid.
The long-dashed and medium-dashed  
lines are for viscous fluid with ADS/CFT ($\eta/s$=0.08) and perturbative ($\eta/s$=0.135) estimate of viscosity. 
Non-equilibrium correction to equilibrium distribution function
is included.}
\label{F6}
\end{figure}

In Fig.\ref{F6}, we have shown the transverse momentum distribution of 
pions obtained in the Cooper-Frye formalism. Freeze-out temperature is $T_F$=0.158 GeV. In this calculation resonance contribution
 to pion spectra is neglected. Pion production is increased in viscous dynamics.   We also note that effect of viscosity
is more prominent at large $p_T$ than at low $p_T$. $p_T$ spectra of pions are flattened with viscosity. Particle production
increases if viscosity
increases. Thus while with ADS/CFT estimate of viscosity,
$\eta/s$=0.08, at   $p_T$ =3 GeV, pion production is increased by a factor 3,
with the perturbative estimate of viscosity, $\eta/s$=0.135, the production
is increased by a factor of 5. Increase is even more at
larger $p_T$.

\begin{figure}
\includegraphics[bb=14 13 581 829,width=0.99\linewidth,clip]{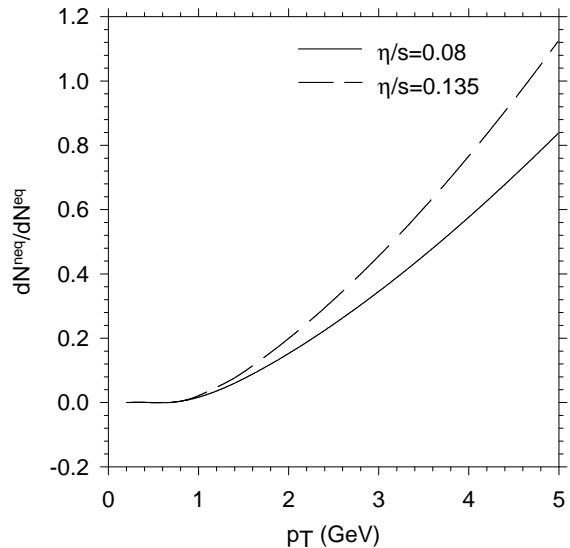}
\vspace{-4.5cm}
\caption{Ratio of correction to particle production due to  non-equilibrium
distribution to equilibrium distribution function.}  
\label{F7}
\end{figure}

We have obtained the non-equilibrium distribution as a correction to the
equilibrium distribution function. It is implied that non-equilibrium effects
are small and  the ratio

\begin{equation}
\frac{dN^{neq}}{dN^{eq}}=\frac
{\frac{dN^{neq}}{dyd^2p_T}}{\frac{dN^{eq}}{dyd^2p_T}},
\end{equation}

\noindent is less than 1. 
In Fig.\ref{F7}, the ratio 
is shown as a function of $p_T$. With ADS/CFT estimate of viscosity, 
$\eta/s$=0.08, non-equilibrium correction to particle 
production become comparable to equilibrium contribution beyond 
$p_T$=5 GeV. However, with perturbative estimate, $\eta/s$=0.135,
non-equilibrium correction become comparable to or exceeds the equilibrium contribution at $p_T$=4.5 GeV. Thus with perturbative
estimate of viscosity, hydrodynamic description
break down above $p_T \sim 4.5 GeV$. 
The blast wave model analysis \cite{Teaney:2004qa} on the otherhand
indicated that viscous dynamics get invalidated beyond 
$p_T \sim$1.7 GeV. The results are not contradictory. In the blast wave model, at the freeze-out,
viscosity is quite large, sound attenuation length $\Gamma_s\sim$ 1.4 fm.
In the present simulation, even for perturbative estimate of viscosity,
sound attenuation length at the freeze-out is $\Gamma_s \sim 0.2 fm$,
7 times smaller than the sound attenuation length used in the blast wave analysis. Naturally, non-equilibrium corrections to
equilibrium distribution
function remains small over an extended $p_T$ range.

\begin{figure}
\includegraphics[bb=14 13 581 829,width=0.99\linewidth,clip]{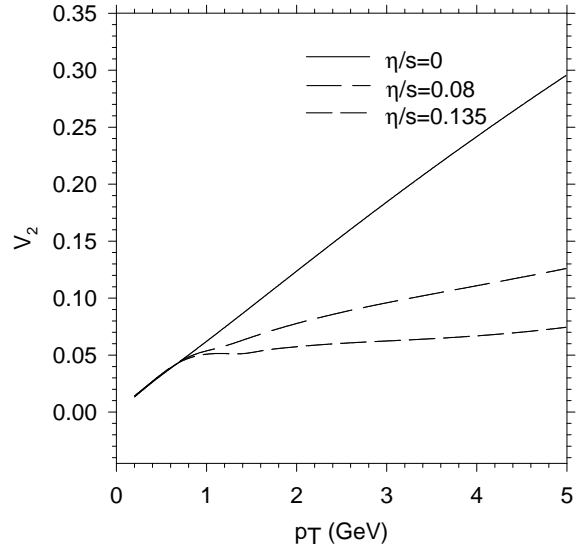}
\vspace{-4.5cm}
\caption{Elliptic flow as a function of transverse momentum. The solid
line is for ideal fluid. The long-dashed, medium-dashed and short-dashed lines are for
viscous fluid with $\eta/s$=0.08 and 0.135 respectively.
Non-equilibrium correction to equilibrium distribution function is included.}
\label{F8}
\end{figure}

We have also calculated the elliptic flow in the model. Being a ratio, 
elliptic flow is very sensitive to the model. Experimentally, elliptic flow
saturates at large $p_T$. It is known
that ideal fluid does not explain the saturation of elliptic flow. 
In contrast to experiment, with ideal fluid, elliptic flow continues to increase
with $p_T$. In Fig.\ref{F8}, we have compared the elliptic flow in ideal and viscous
fluid.
The solid line is $v_2$ for the ideal fluid. The long-dashed and medium-dashed 
lines are  for viscous fluid with ADS/CFT ($\eta/s$=0.08) and 
perturbative ($\eta/s$=0.135) estimated viscosity.
Elliptic flow decreases with viscosity. As viscosity increases, elliptic flow
is also reduced. We also note that both for ADS/CFT and perturbative estimate of viscosity, elliptic flow indicate
saturation at large $p_T$.
The result is very encouraging, as experimentally also elliptic flow
tends to saturate at large $p_T$.
 
As discussed earlier, ideal fluid dynamics,   can explain a large volume of data in Au+Au 
collisions at RHIC.  Our present knowledge about the hot dense matter
produced in central Au+Au collisions are obtained from the ideal
fluid analysis.
As shown in the present paper, QGP fluid,  even with 
ADS/CFT estimate of viscosity $\eta/s$=0.08, generate enough entropy to enhance particle production by a factor of 3 at $p_T$=3
GeV. Naturally,
if QGP fluid is viscous,  initial conditions
as required to explain RHIC data with ideal fluid dynamics will overpredict
the experimental $p_T$ distribution. Viscous fluid dynamics will require
much less initial temperature than an ideal fluid to explain the same
$p_T$ spectra.
 As an example, in Fig.10, we have compared the pion spectra  obtained 
in  viscous dynamics with ADS/CFT estimate of
viscosity ($\eta/s$=0.08), initialised with  entropy density of 110,80,
and 60 $fm^{-3}$ with the pion spectra obtained in an 
ideal fluid dynamics, initialised with entropy density s=110 $fm^{-3}$. 
For all the fluids, the initial time is $\tau_i$=0.6 fm and freeze-out temperature
is 158 MeV.
Viscous fluid initialised with entropy density between 60-80 $fm^{-3}$,
 compare well with the pion spectra from ideal fluid initialised
at much higher entropy density. 
To produce the same pion spectra, while ideal
fluid require a initial temperature of 350 MeV, the viscous fluid require
much less temperature between 270-290 MeV. The ideal fluid dynamics
can overestimate the initial temperature of fluid produced in
Au+Au collisions at RHIC by 20-30\%.  

\begin{figure}
\includegraphics[bb=14 13 581 829,width=0.99\linewidth,clip]{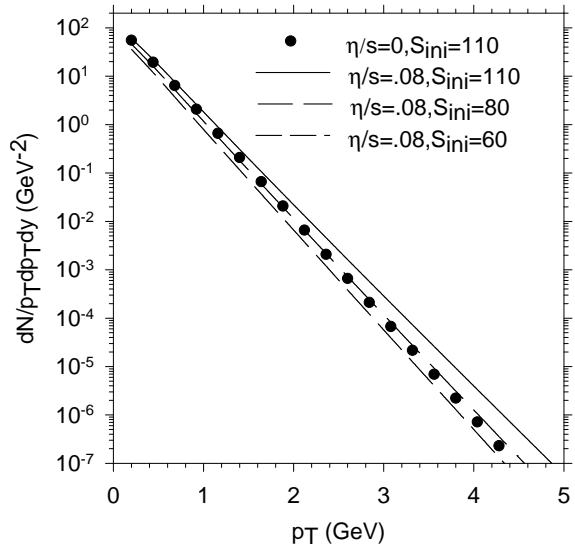}
\vspace{-4.5cm}
\caption{The solid circle is the $p_T$ distribution obtained 
in ideal fluid dynamics, with initial entropy density s=110$ fm^{-3}$. 
The solid, long dashed and dashed lines are for viscous fluid with ADS/CFT estimate of viscosity, $\eta/s$=0.08, initialised
at entropy density
s=60,80 and 110 $fm^{-3}$ respectively. Non-equilibrium correction to equilibrium distribution function is included.}
\label{F10}
\end{figure}
 
\section{Summary and conclusions}

We have studied the boost-invariant hydrodynamic evolution of 
QGP fluid with dissipation due to shear viscosity. In this study we 
have employed the 1st order theory of dissipative relativistic fluid. 
1st order theories suffer from the problem of causality, signal can 
travel faster than light.  Unphysical effects like
reheating of the fluid, early in the evolution, can occur. However, for
a fluid like QGP, where viscosity is small, with appropriate initial
conditions, effects of causality violation can be minimised.
In this model study, we have 
considered two values of viscosity, the ADS/CFT motivated value,
$\eta/s \approx$0.08 and perturbatively estimated viscosity, $\eta/s \approx$0.135.
Both the ideal and viscous fluids are initialised similarly. At the initial 
time $\tau_i$=0.6 fm, initial central entropy density is 110 $fm^{-3}$, 
with transverse profile taken from a Glauber model calculation. 
Viscous hydrodynamics require initial conditions  for the shear-stress tensor components. It is assumed that at the equilibration
time, the shear stress
tensors components have reached their boost-invariant values. 
The initial conditions of
the fluid are such that for both the values of viscosity ($\eta/s$=0.08 and 0.135), the condition of validity of viscous
hydrodynamics, $\Gamma_s/\tau << 1$ is satisfied all through the evolution.
Explicit simulation of ideal and viscous fluids confirms that energy density of a 
viscous fluid, evolve slowly than its ideal counterpart. The fluid velocities
on the other hand evolve faster in viscous dynamics than in ideal 
dynamics. Transverse expansion is also more in viscous dynamics.
For a similar freeze-out condition freeze-out surface is extended in
viscous fluid.

We have also studied the effect of viscosity on particle production.
Viscosity generates entropy leading to enhanced particle production.
Particle production is  increased due to (i) extended freeze-out surface and (ii) non-equilibrium
correction to equilibrium distribution function. Non-equilibrium
correction to equilibrium distribution function is a dominating
factor influencing the particle production
at large $p_T$.
  With ADS/CFT (perturbative) estimate of viscosity, at $p_T$=3 GeV,
pion production is increased by a factor 3 (5) . Increase is even more at large $p_T$. While viscosity enhances particle
production,
it reduces the elliptic flow. At $p_T$=3 GeV, for ADS/CFT(perturbative) estimate of viscosity,
elliptic flow is reduced by a factor of 2(3). We also find that at large $p_T$
elliptic flow tends to saturate.

To conclude, present study shows viscosity, even if small, can be
very important in analysis of RHIC Au+Au collisions. Currently accepted
initial temperature of hot dense matter produced in RHIC Au+Au collisions,
obtained from ideal fluid analysis
can be changed by 20\% or more with dissipative dynamics. 
 
\acknowledgements
Prof. U. Heinz initiated this programme of numerical 
simulation of dissipative hydrodynamics in 2+1 dimension.  
The author would like to thank Prof.  Heinz for several discussions
and suggestion.

\end{document}